\documentclass[a4paper,aps,pre,reprint,superscriptaddress,showpacs]{revtex4-1}

\usepackage[latin1]{inputenc}    
\usepackage{amsmath}           
\usepackage{amsfonts}          
\usepackage{amssymb}           
\usepackage{lmodern,dsfont}	
\usepackage{color}
\usepackage{graphicx}       
\usepackage[T1]{fontenc}    
\begin{document}

\title{On the effect of the drive on self-organized criticality}
\author{Marco Winkler}
\email{mwinkler@physik.uni-wuerzburg.de}
\author{Johannes Falk}
\email{jfalk@physik.uni-wuerzburg.de}
\author{Wolfgang Kinzel}
\email{kinzel@physik.uni-wuerzburg.de}
\affiliation{Universit\"at W\"urzburg, Fakult\"at f\"ur  Physik und Astronomie, 97074  W\"urzburg, Germany}

\def\d{{\rm d}}
\def\0{\emptyset}

\def\comment#1{\color{red}[\textbf{comment: #1}]\color{black}}
\def\mark#1{\color{red}#1 \color{black}}

\begin{abstract}

The well known Sandpile model of self-organized criticality generates avalanches of all length and time scales, without tuning any parameters. In the original models the external drive is randomly selected. Here, we investigate a drive which depends on the present state of the system, namely the effect of favoring sites with a certain height in the deposition process.
If sites with height three are favored, the system stays in a critical state. Our numerical results indicate the same universality class as the original model with random deposition, although the stationary state is approached very differently. In contrast, when favoring sites with height two, only avalanches which cover the entire system occur. Furthermore, we investigate the distributions of sites with a certain height, as well as the transient processes of the different variants of the external drive.

\end{abstract}

\maketitle
\parskip 1mm

\section{Introduction}

Systems far from equilibrium can reach a state where avalanche-like events occur on a broad distribution of length and time scales.
Some systems even seem to be scale free: The distribution of events follows a power law over several orders of magnitude. Examples are found in the context of earthquakes, sandpiles, neural networks, synchronization processes, evolution, superconductors, magnets, turbulence, ecology, financial markets, data networks and more \cite{b._gutenberg_seismicity_1949, newman2005power, price_networks_1965,Winkler2012}. In thermal equilibrium, 
scale-free correlations exist for continuous phase transitions at a critical point (temperature, pressure, etc.). Therefore, scale-free phenomena far from equilibrium are called \emph{critical}. In case the critical behavior is generated from the dynamics of interacting units, this phenomenon is coined \emph{self-organized criticality (SOC)}.

In 1987, Bak, Tang, and Wiesenfeld (BTW) introduced a simple model which shows SOC \cite{bak_self-organized_1987}. Since then, in about 5000 publications, this model, its variants, and possible applications have been investigated in great detail \cite{dhar_abelian_1999, dhar_theoretical_2006, christensen_sandpile_1993, pruessner2012selfcontrol,dupoyet_replicating_2010, paoletti_deterministic_2013}. The model consists of discrete or continuous variables on a lattice and contains three essential mechanisms: a drive, a sequence of topplings and a separation of time scales. The drive increases one of the local variables. If this variable exceeds a threshold value, it is distributed among its neighbors. As a consequence, the neighbors may also be shifted above threshold and continue to distribute, hence generating an avalanche. Only after the avalanche has stopped, a new drive is activated.

Numerous variants of the BTW model have been introduced and investigated. In particular, several mechanisms of topplings -- deterministic, stochastic, conservative or dissipative \cite{chessa_universality_1999,malcai_dissipative_2006} -- have been studied, and the corresponding cooperative behavior has been calculated, mainly numerically \cite{lubeck_numerical_1997}. However, although there exist  models that are inherently defined by a state-dependent drive~\cite{bak1993punctuated,olami1992self}, to our knowledge, the explicit effect of different driving mechanisms has not been investigated so far. Usually, a site is selected randomly and its local variable is increased by one.

In this paper we study the effect of a state-dependent drive. A site is still chosen randomly, but now the local variable is increased only if it has some specific value. 
The drive selects specific values of height, energy, charge or slope, depending on possible applications in mind. Only if variables with this value do not exist, other sites are driven randomly. We investigate, whether SOC is destroyed by this mechanism, or to what extend the critical properties are modified. 

The remainder of this article is organized as follows. In section~\ref{sec:model} we will recapitulate the definition of the regular Sandpile model and introduce our modified versions of it. In section~\ref{sec:results} we will present avalanche-size distributions of the models and take a detailed look at the models' time evolutions both in the steady state and in the transient phase.

\section{Model} \label{sec:model}

\subsection{Sandpile model}

We consider the model of Dhar which has the structure of an Abelian group ~\cite{dhar_self-organized_1990}. Let us briefly review its definition. The model is defined on a finite two-dimensional square lattice, where each site has a value representing the height of the pile. We denote the height of each site at position $x$ and $y$ with $h(x,y)$. In the original model the initial configuration is an empty field, where for every $x$ and $y$: $h(x,y) = 0$. The process evolves by random-sequential updates putting grains on the field according to the following rule:
\begin{equation}
h(x,y) \to h(x,y)+1
\label{eq:drive}
\end{equation}
As soon as one site grows taller than three it gets unstable and topples, i.e. four of its grains are distributed equally among its four neighbors:
\begin{equation}
 \begin{split}
h(x,y) &\to h(x,y)-4\\
h(x \pm 1, y) &\to h(x \pm 1, y) + 1 \\
h(x, y \pm 1) &\to h(x, y \pm 1) + 1 
 \end{split}
\label{eq:avalanches}
\end{equation}
The boundary conditions are chosen in such a way, that all grains that topple out of the system are dissipated. This mechanism can create avalanches on all scales that can reach over the full grid.

It can be shown, that the final configuration is independent on the order of the sites that are toppled. For this reason the model is often called Abelian Sandpile model \cite{dhar_self-organized_1990}.

In the literature, different methods of measuring the size of the avalanches exist. The two most widely used are:
\begin{description}
\item[number of topplings] Here, the number of topplings that occur in one avalanche are counted.
\item[number of involved sites] Here, the number of involved sites are counted. In difference to the first method each site can only be counted once, even if it topples twice or more.
\end{description}
In this paper, we focus on the second method of counting the involved sites. 

Without any tuning of parameters this model is attracted to its critical state where various quantities $q$ of the system are power-law distributed: 
\begin{align}
P(q) \simeq q^{-\alpha}
\end{align}
Here, $P(q)$ is the probability density function of $q$ and $\alpha$ is the critical exponent, that can be used to classify the universality class. 
It was shown that the critical exponent for the distribution of the avalanche sizes is close to one, $\alpha \simeq 1$. For a summary of numerical results see e.g.~\cite[Table 4.1]{pruessner2012selfcontrol}. It is also well known, that the average height, $\left\langle h \right\rangle$, of the critical system fluctuates around $\left\langle h \right\rangle = 17/8  \simeq 2.1$ \cite{jeng_height_2006, priezzhev_structure_1994}. 

\subsection{Modified Sandpile model with state-dependent drive}

In order to investigate the implications of a state-dependent drive on the model, we slightly modify the update rules~\eqref{eq:drive} and~\eqref{eq:avalanches}. For our modified model, we change the random-sequential update to a favor-$n$ random-sequential update, where $n \in \left\{0,1,2,3\right\}$. This means for a favor-$n$ update:
\begin{description}
\item[if at least one site has height $n$] put the next grain on a randomly chosen site of height $n$
\item[if no site hast height $n$] put the next grain on any randomly chosen site.
\end{description}
These modified update rules, as well as the original Sandpile model, can be interpreted as special cases of a more general model in which particles are added at randomly selected sites and reflected with probabilities ($p_0$,$p_1$,$p_2$,$p_3$) depending on the current height at the chosen site. Then, in the favor-$n$ model all sites reflect a new particle with probability one unless they have height $n$. In the latter case they accept the particle with probability one, i.e. $p_i = (1-\delta_{in})$. Contrary, the original Sandpile model will accept particles at any height, $p_i = 0$.

Besides the mentioned modifications, the triggering of avalanches still proceeds according to the update rules in Eq.~\eqref{eq:avalanches}.

\section{Results} \label{sec:results}
We will now present simulation results for the modified Sandpile model. In our numerical analysis we compare the distributions of avalanches in the steady state for the unmodified model and the modified one. Furthermore, we focus on the behavior of the systems in the transient phase from the empty lattice to the steady state.

\subsection{Avalanche Distributions}

\begin{figure}
\centering\includegraphics[width=0.5\textwidth]{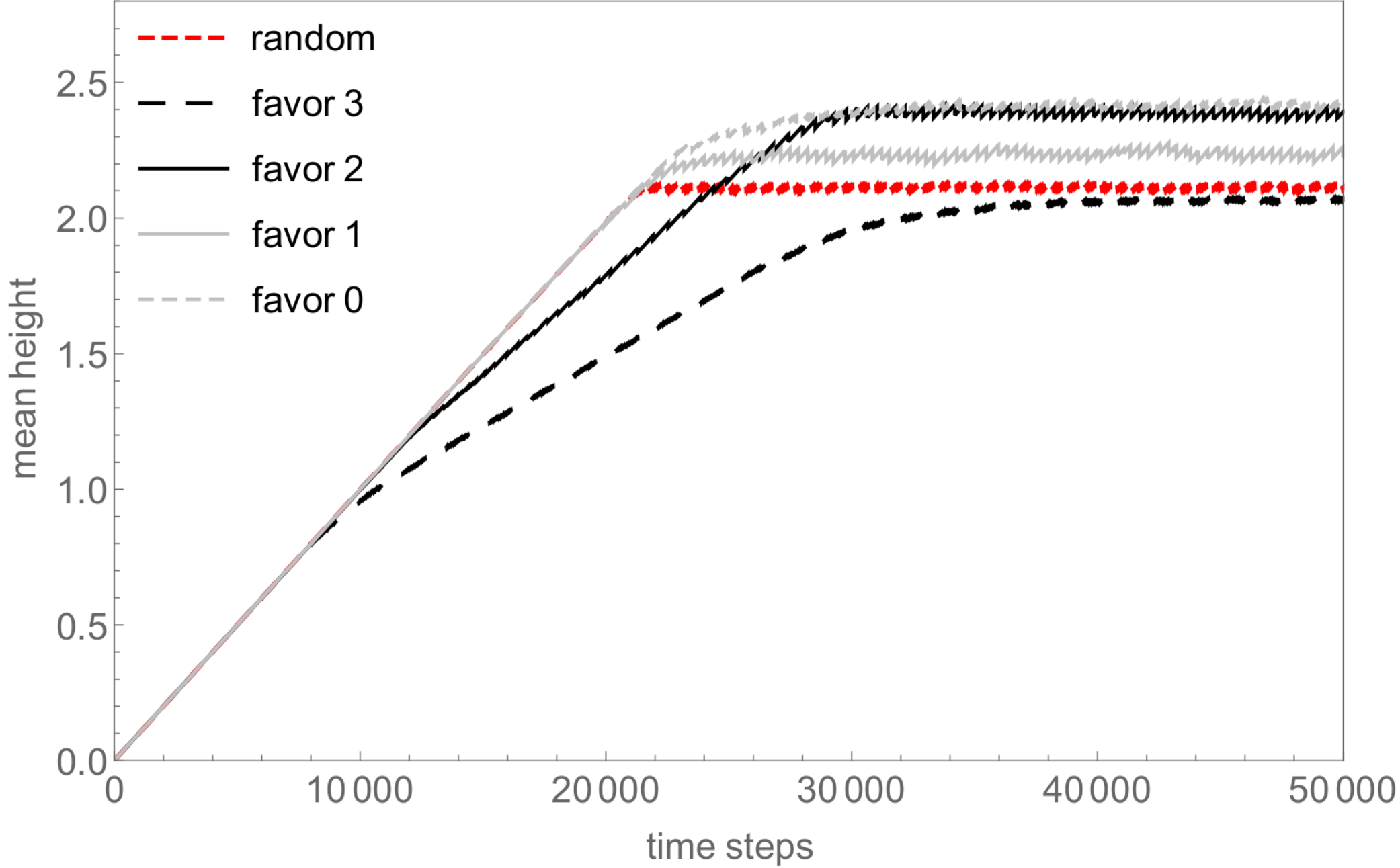}
\caption{(Color online) Transient of the mean system height for the unmodified Sandpile model (with random deposition) and different favor-$n$-models.}
\label{fig:mean_system_height}
\end{figure}

To estimate whether the system already reached the steady state, we measure the evolution of the mean system height $\left\langle h \right\rangle$, i.e. the average occupation per site. Fig.~\ref{fig:mean_system_height} shows that independently on the chosen model, after about $4.0 \times 10^4$ time steps the system is in a steady state. Some favor-$n$-models show strong oscillations in their steady state indicating large-scale avalanches.

\begin{figure}
\centering\includegraphics[width=0.5\textwidth]
{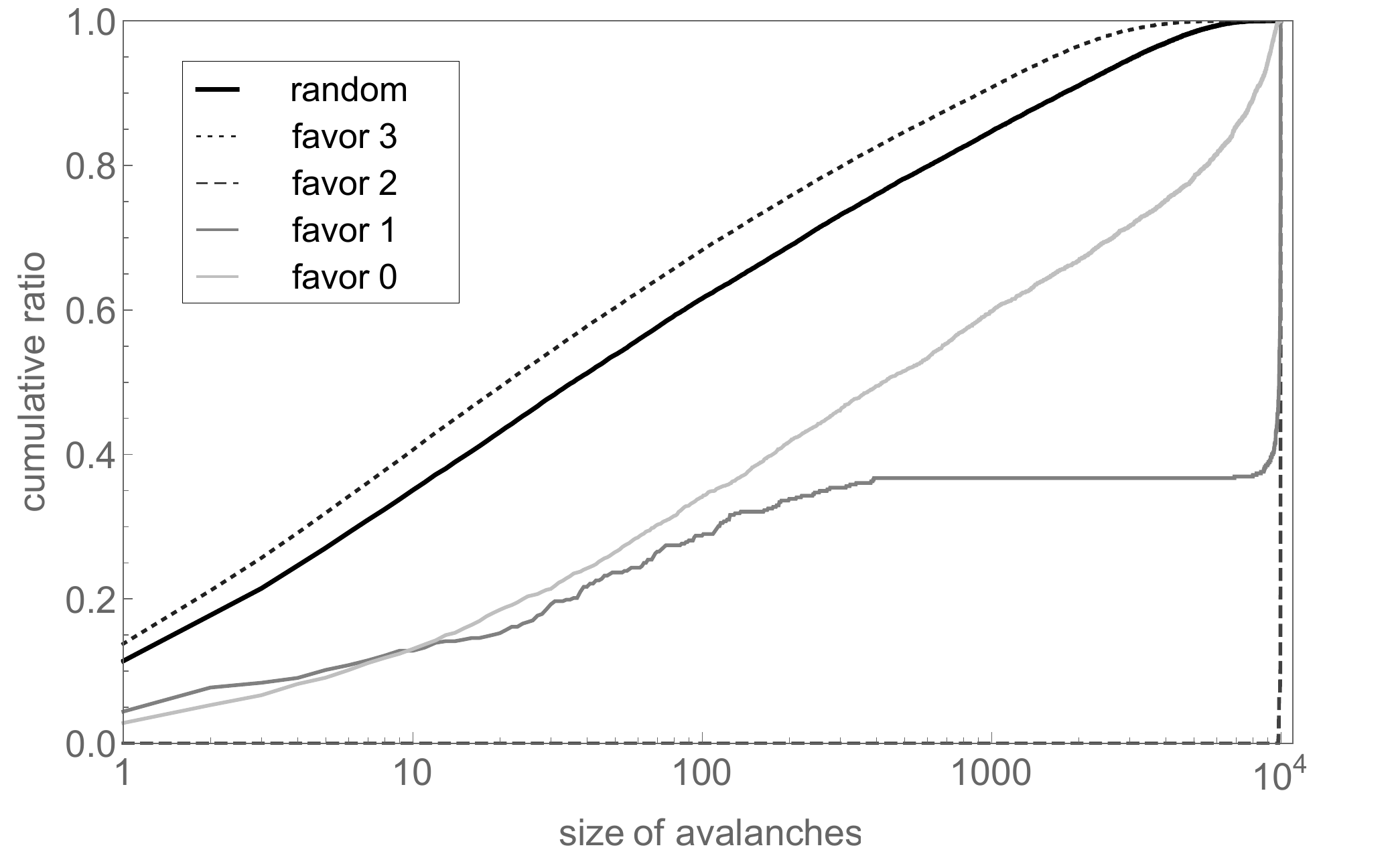}
\caption{Cumulative distribution of the avalanche sizes in a system of size $100 \times 100$. Note the linear-logarithmic scale.}
\label{fig:avalanche_size}
\centering\includegraphics[width=0.5\textwidth]{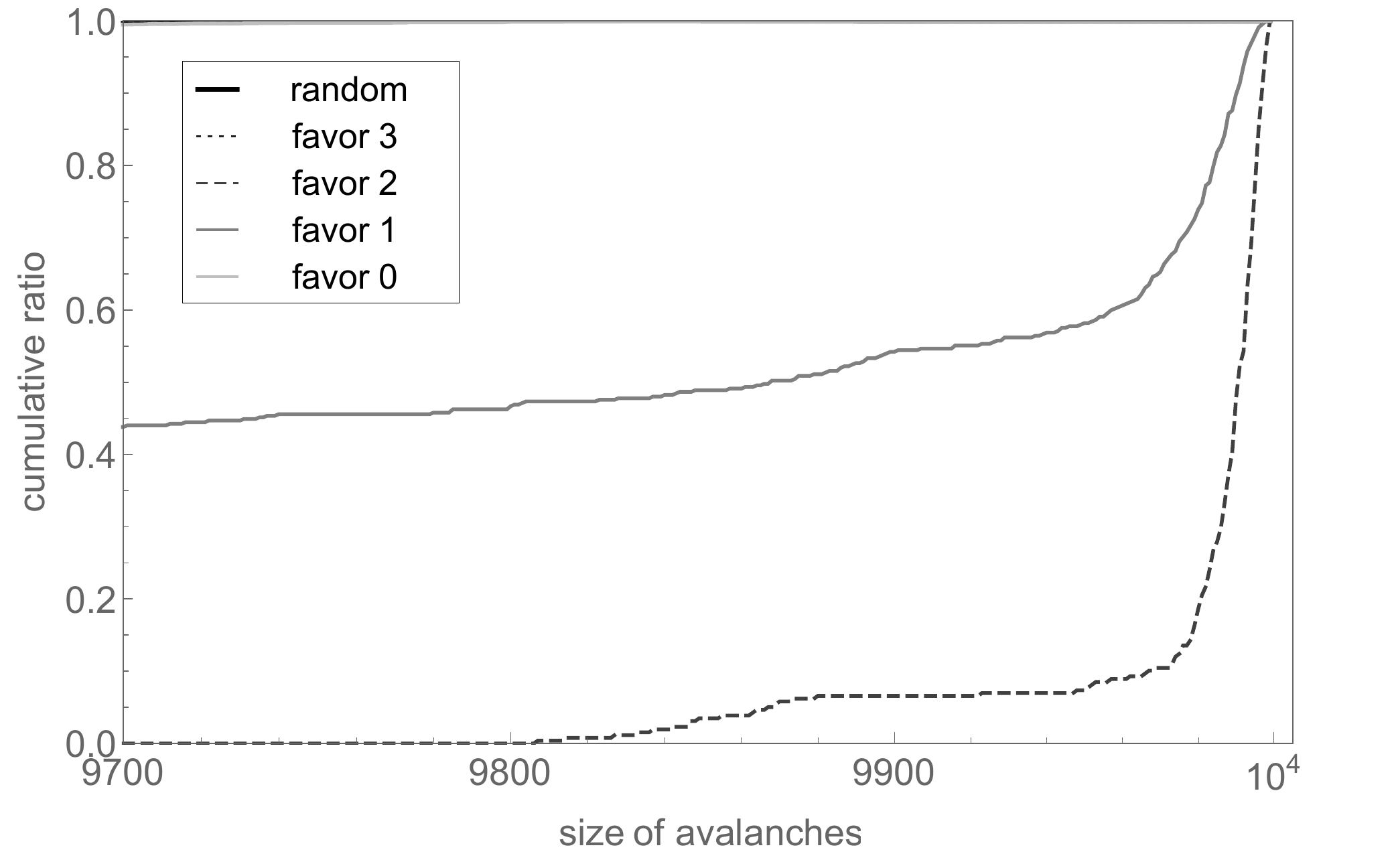}
\caption{Focused view at the end of the cumulative distribution of the avalanche sizes in a system of size $100 \times 100$.}
\label{fig:avalanche_size_focus}
\end{figure}

The cumulative distribution of avalanche sizes is shown in Fig.~\ref{fig:avalanche_size}. We find that the favor-3 model stays in a critical state. Even the exponent of the original model and the favor-$3$ model seem to be the same, namely
\begin{equation}
\alpha_\text{random} \simeq \alpha_\text{favor3} \simeq 1.021 \pm 0.033
\label{eq:critExpRnd3}
\end{equation}
This is one of the main results of this paper: The critical properties are not changed although the mechanism of the drive is very different. 

Contrary, for the favor-$1$ and the favor-$0$ model the distributions seem to be non-critical. In addition to some small-scale avalanches, the favor-1 model has a peak around the system size of $10^4$ (for a detailed view at the very end of the distribution see Fig.~\ref{fig:avalanche_size_focus}). The favor-0 model shows avalanches on various scales, yet the distribution seems to be non-critical as well. The favor-$2$ model cannot be seen at all in Fig.~\ref{fig:avalanche_size}. In Fig.~\ref{fig:avalanche_size_focus} we see that this model only creates avalanches that approximately cover the entire system.

Hence, there exist also mechanisms of the drive, which change the properties of the final stationary state. There is still a distribution of avalanches, but - depending on the drive - the system favors avalanches which occupy the complete lattice.

\subsection{Site occupations}

\begin{figure}
\centering\includegraphics[width=0.5\textwidth]
{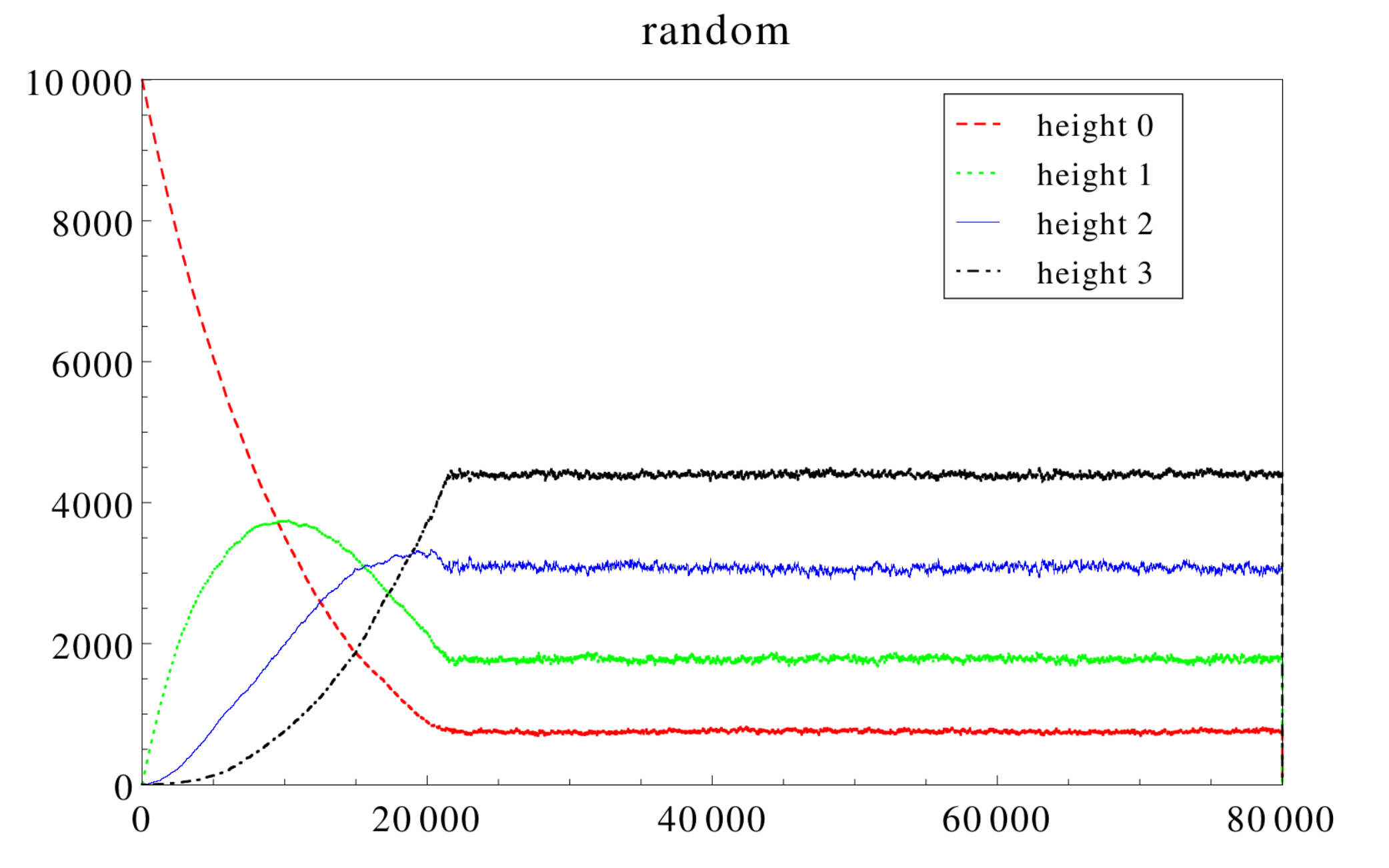}
\caption{(Color online) Distribution of the different site heights in a system of size $100 \times 100$ doing random deposition.}
\label{fig:height_dist_rand}
\end{figure}

To better understand the behavior for different drives, we analyze the distribution of distinct site heights depending on the model chosen. Fig.~\ref{fig:height_dist_rand} shows the results for the unmodified Sandpile model with random deposition. Its site-occupation distribution in the steady state is known analytically~\cite{jeng_height_2006}:
\begin{equation}
\begin{split}
	p\left(0\right) &= \frac{2}{\pi^2} - \frac{4}{\pi^3} \simeq 7.4\%\\
	p\left(1\right) &= \frac{1}{4} - \frac{1}{2 \pi} - \frac{3}{\pi^2} + \frac{12}{\pi^3} \simeq 17.4\%\\
	p\left(2\right) &= \frac{3}{8} + \frac{1}{\pi} - \frac{12}{\pi^3} \simeq 30.6\%\\
	p\left(3\right) &= \frac{3}{8} - \frac{1}{2 \pi} + \frac{1}{\pi^2} + \frac{4}{\pi^3} \simeq 44.6\%
\label{eq:siteOccRnd}
\end{split}
\end{equation}

\begin{figure}
\centering\includegraphics[width=0.5\textwidth]{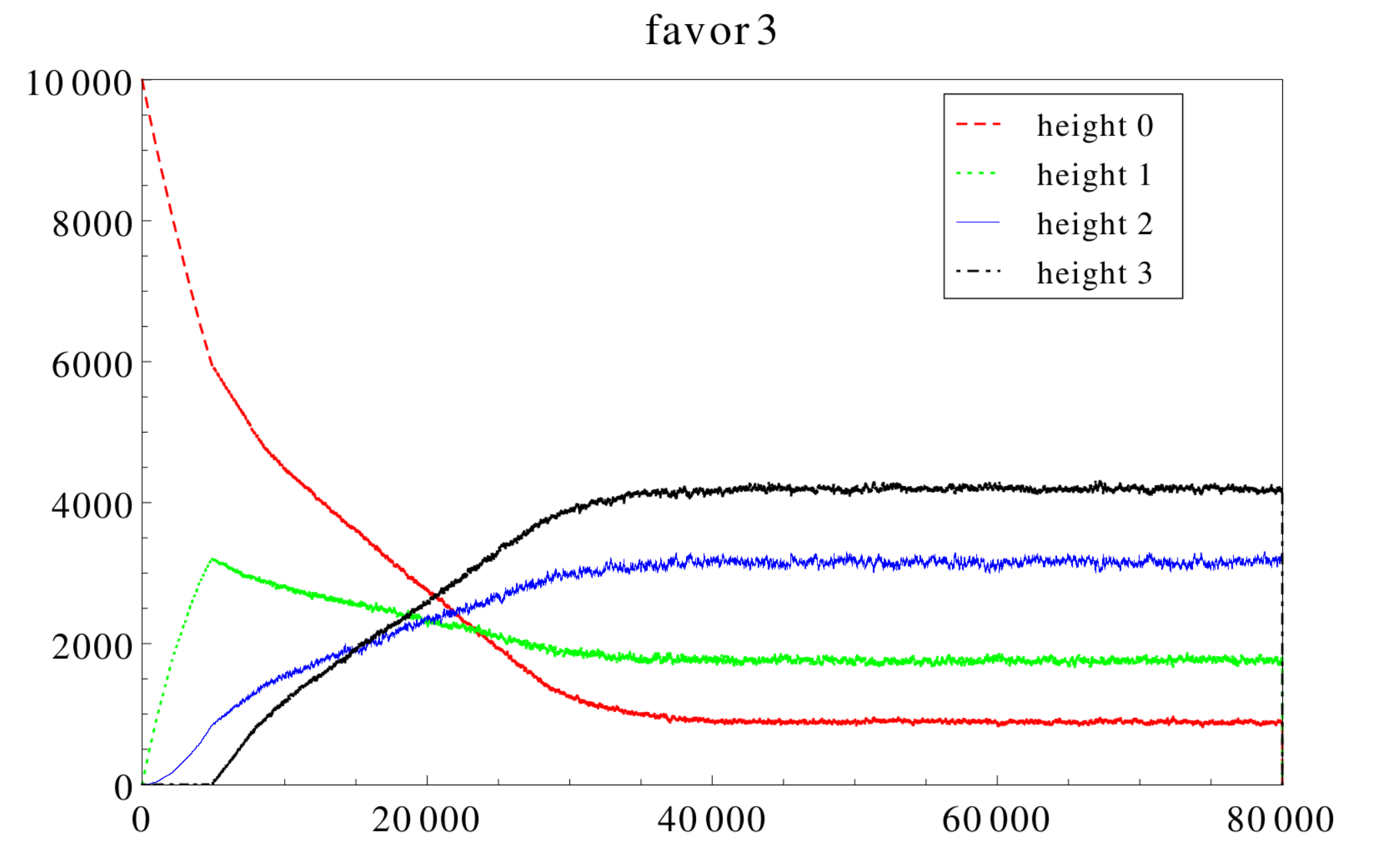}
\caption{(Color online) Distribution of the different site heights in a system of size $100 \times 100$ doing favor-$3$ deposition}
\label{fig:height_dist_3}
\end{figure}

\begin{figure}
\centering\includegraphics[width=0.5\textwidth]{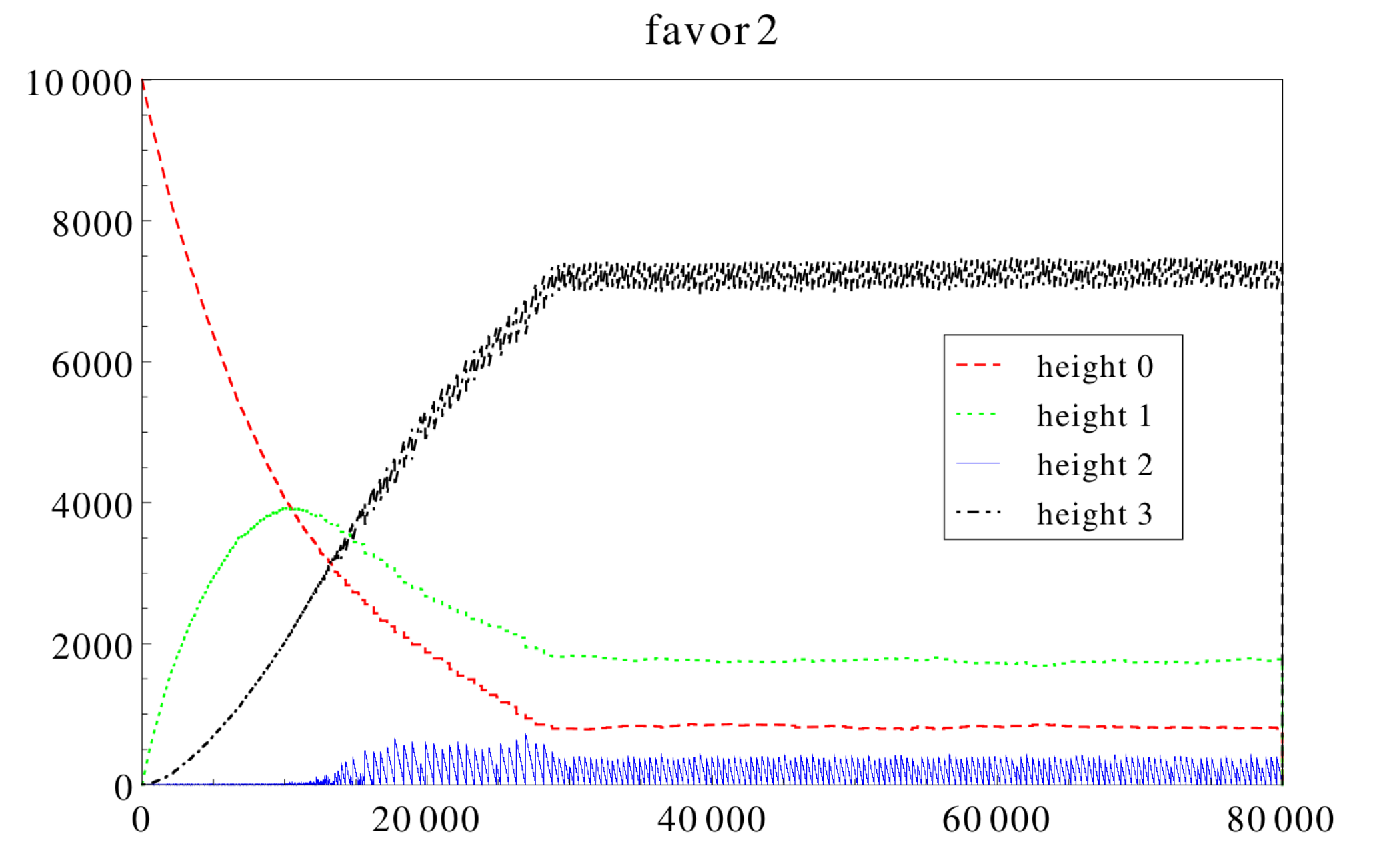}
\caption{(Color online) Distribution of the different site heights in a system of size $100 \times 100$ doing favor-$2$ deposition}
\label{fig:height_dist_2}
\end{figure}

\begin{figure}
\centering\includegraphics[width=0.5\textwidth]
{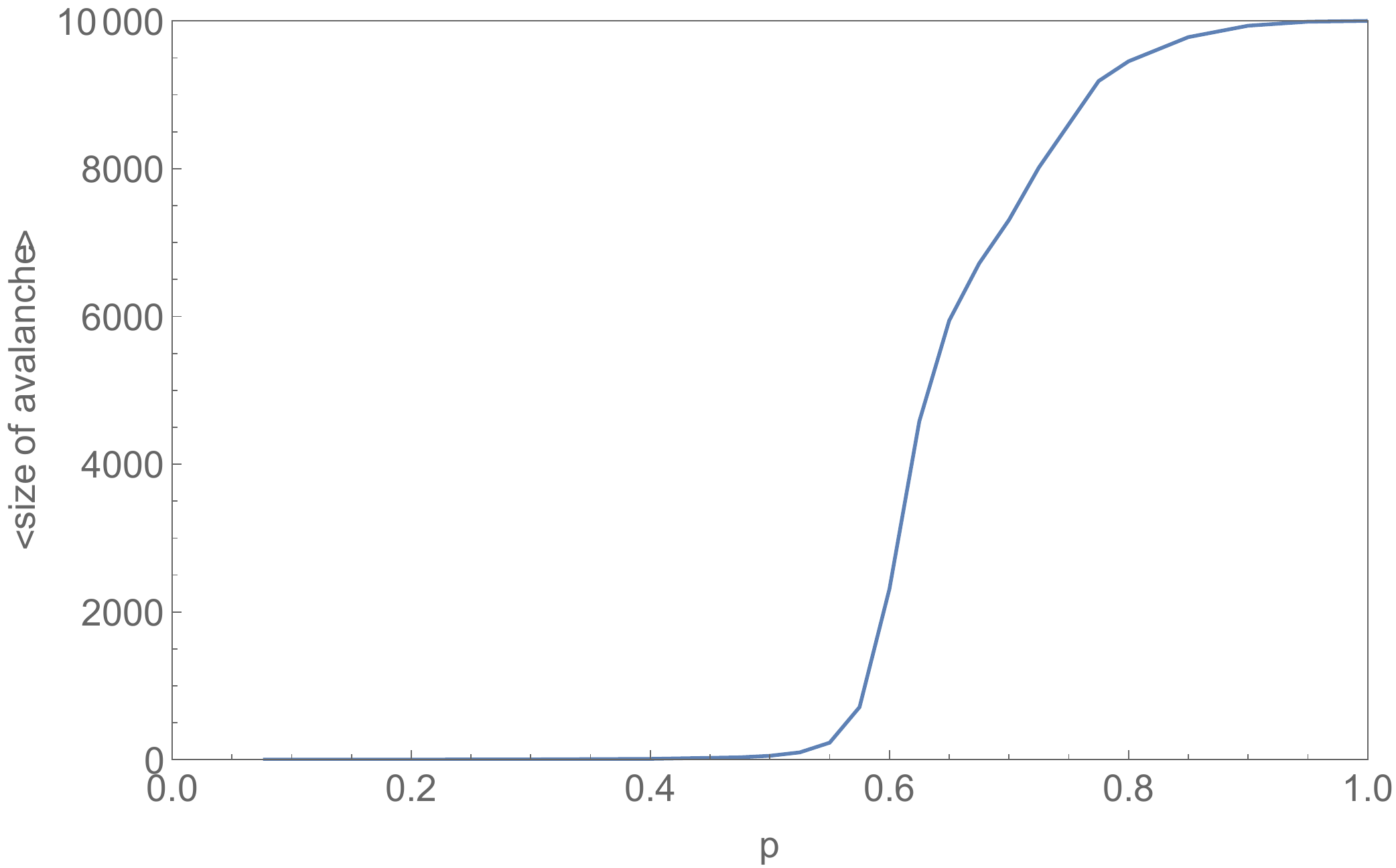}
\caption{Average size of triggered avalanches in a system of size $100 \times 100$ where site heights are randomly set to 3 with probability $p$ and to height $0$ or $1$ with probability $(1-p)/2$, respectively. Averaged over $5 \times 10^5$ runs for each $p$.}
\label{fig:phase_03}
\end{figure}

\begin{figure}
\centering\includegraphics[width=0.5\textwidth]{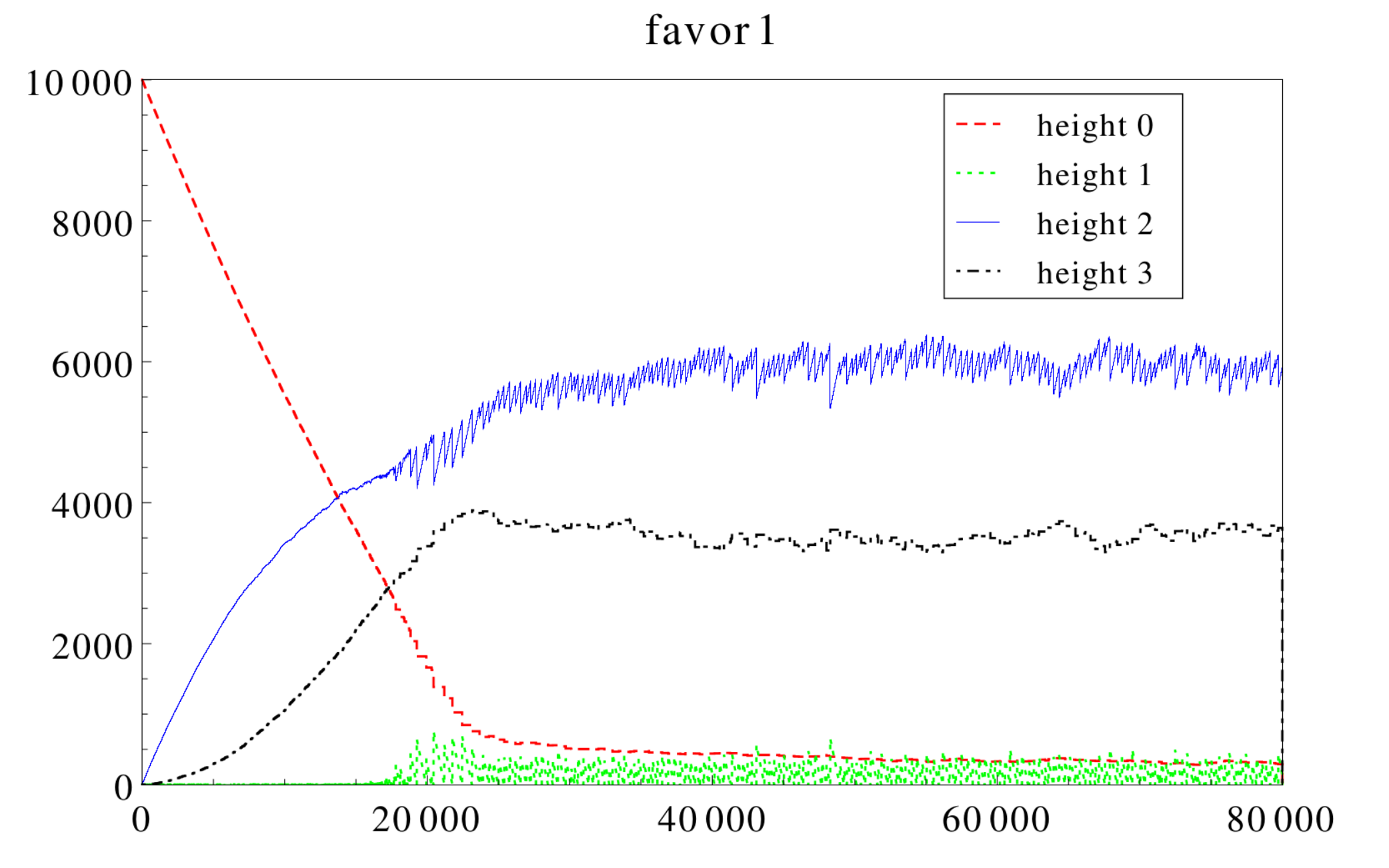}
\caption{(Color online) Distribution of the different site heights in a system of size $100 \times 100$ doing favor-$1$ deposition}
\label{fig:height_dist_1}
\end{figure}

The distribution in the steady state of the favor-$3$ model (see Fig.~\ref{fig:height_dist_3}) is very similar to the one of the random-deposition model (Fig.~\ref{fig:height_dist_rand}):
\begin{equation}
\begin{split}
	p\left(0\right) &\simeq 9.1\%\\
	p\left(1\right) &\simeq 17.2\%\\
	p\left(2\right) &\simeq 31.7\%\\
	p\left(3\right) &\simeq 42.0\%
\label{eq:siteOccf3}
\end{split}
\end{equation}

This already indicates that the two models, random and favor-3, have the same properties in the stationary state.

However, the transient properties of the two models are different. The approach to the final state differs already in the mean occupation numbers as a function of time steps, and the curves of the favor-3 model show kinks, whereas the ones of the Sandpile model are smooth (Figs.~\ref{fig:height_dist_rand} and~\ref{fig:height_dist_3}). 
The kinks in the favor-3 case coincide with the appearance of the first site with height three, and hence the first avalanche, which makes it difficult to model. The different transient behavior between the models will be further discussed in subsection~\ref{sec:transient}. Moreover, the final transition between the transient and the steady state causes kinks in the random model while the favor-3 model converges slowly to the stationary distribution.

\begin{figure}
\centering\includegraphics[width=0.5\textwidth]
{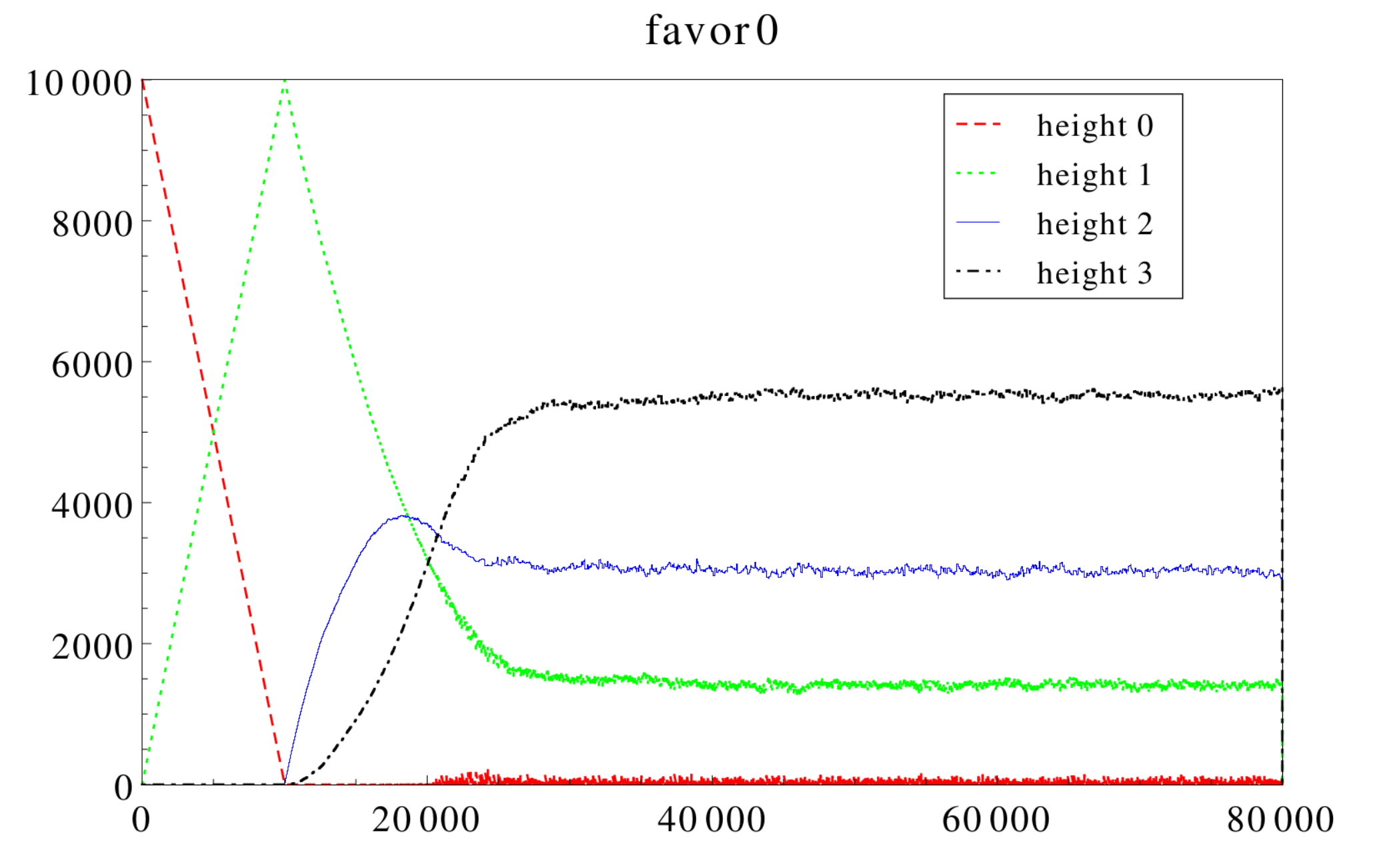}
\caption{(Color online) Distribution of the different site heights in a system of size $100 \times 100$ doing favor-$0$ deposition}
\label{fig:height_dist_0}
\end{figure}

In contrast to the random and the favor-3 case the occupation numbers for the other favor-$n$ models look -- even qualitatively -- very different (Fig.~\ref{fig:height_dist_2}, Fig.~\ref{fig:height_dist_1}, and Fig.~\ref{fig:height_dist_0}).

For the favor-$2$ model we already found from Fig.~\ref{fig:avalanche_size} and Fig.~\ref{fig:avalanche_size_focus} that the avalanches cover the whole system. In the steady state of this model, sites of height two are immediately grown to sites of height three and are therefore supressed in the distribution of Fig.~\ref{fig:height_dist_2}. We find the following occupation numbers for the favor-2 drive:
\begin{equation}
\begin{split}
	p\left(0\right) &\simeq 8.9\%\\
	p\left(1\right) &\simeq 17.7\%\\
	p\left(2\right) &\simeq 0.0\%\\
	p\left(3\right) &\simeq 73.4\%
\label{eq:siteOccf2}
\end{split}
\end{equation}
To understand how these site occupations may be related to the fact that all avalanches cover basically the whole system, let us consider a system prepared in the following way. Sites are randomly set to a height of  $3$ with probability $p$ and a height of $0$ or $1$ with probability $(1-p)/2$, respectively. Putting now a new grain on a site of height $3$ launches an avalanche. Of course, the expected size of this avalanche depends on the distribution of fields with height $3$, $1$, and $0$, and therefore on $p$. Fig.~\ref{fig:phase_03} shows the average size of triggered avalanches versus $p$. The shape of the curve indicates a phase transition with a critical value of $p_c \simeq 0.58$. Below $p_c$ avalanches cover only a finite region of the system, while above that threshold their size grows to infinity. I.e., if more than $58\%$ of the sites have a height of $3$, the avalanche covers the entire system. Relating this result to the distribution in Fig.~\ref{fig:height_dist_2}, it becomes clear that, in the favor-2 case, all avalanches cover the whole system, since the percentage of sites with height $3$ is clearly above $60\%$.

Note that the percolation threshold for site percolation on a square lattice is $p_0 \simeq 0.592$ \cite{jacobsen_high-precision_2014}. Hence, if the fraction $p$ of randomly chosen sites with height three is above this percolation threshold, an avalanche will cross the complete lattice. If $p$ is somewhat smaller, we cannot exclude that such large avalanches occur, since, over the course of an avalanche, e.g. sites of height two may get added two particles from their neighbors and also topple. Our numerical results show that the critical threshold for avalanches is very close to the one of percolation.

\begin{figure}
\centering\includegraphics[width=0.5\textwidth]
{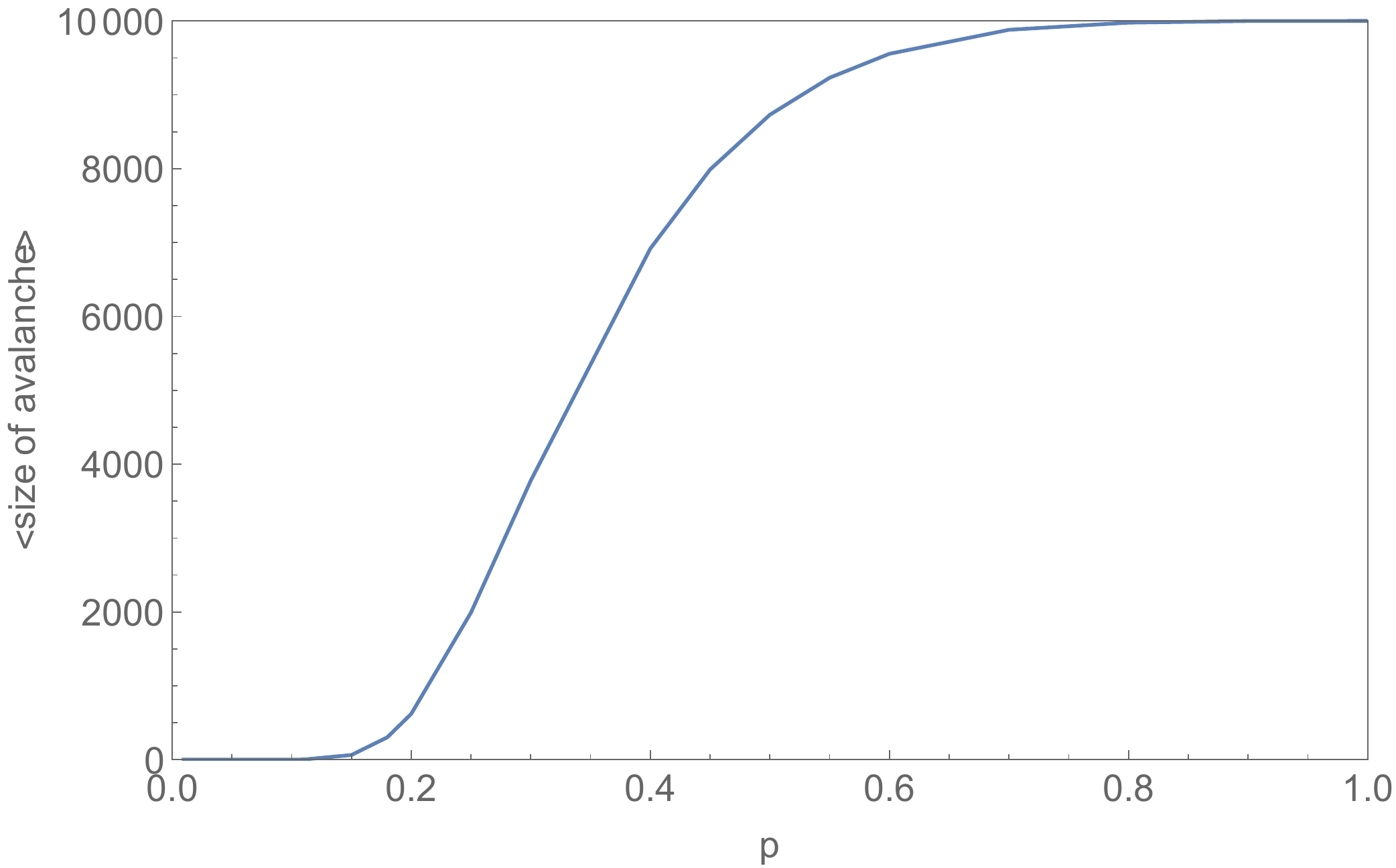}
\caption{Average size of triggered avalanches in a system of size $100 \times 100$ where site heights are randomly set to 3 with probability $p$ and to height $2$ with probability $(1-p)$. Averaged over $5 \times 10^5$ runs for each $p$.}
\label{fig:phase_23}
\end{figure}

In Fig.~\ref{fig:avalanche_size} and Fig.~\ref{fig:avalanche_size_focus}, the distribution of avalanche sizes for the favor-1 model has two parts: a broad distribution of smaller sizes and a peak of avalanches covering the whole system. For the distribution of site occupations in the steady state, we find (see Fig.~\ref{fig:height_dist_1})
\begin{equation}
\begin{split}
	p\left(0\right) &\simeq 2.5\%\\
	p\left(1\right) &\simeq 0.0\%\\
	p\left(2\right) &\simeq 60.8\%\\
	p\left(3\right) &\simeq 36.7\%
\label{eq:siteOccf1}
\end{split}
\end{equation}
Since sites of height one are immediately grown to height two, the former are supressed and hardly appear in the distribution. To understand the appearance of the peak of avalanches spanning the whole system let us prepare a setup, where the sites are randomly distributed with a height of three with probability $p$ and a height of two with probability $1-p$. Fig.~\ref{fig:phase_23} shows, that for this case the critical value of $p$ is about $p_c \simeq 35 \% $. The latter is very close to our observed ratio of fields with height three in Eq.~\eqref{eq:siteOccf1}. Hence, we observe both a distribution of small-scale avalanches and a peak of avalanches spanning the entire system.

The steady-state distribution of sites for the favor-0 model is shown in Fig.~\ref{fig:height_dist_0}:
\begin{equation}
\begin{split}
	p\left(0\right) &\simeq 0.0\%\\
	p\left(1\right) &\simeq 13.9\%\\
	p\left(2\right) &\simeq 30.4\%\\
	p\left(3\right) &\simeq 55.7\%
\label{eq:siteOccf0}
\end{split}
\end{equation}

\subsection{Transient phase} \label{sec:transient}

From our earlier observations, we have learned that, in the steady state, the original Sandpile model and the favor-3 model yield similar distributions of both avalanche sizes (Fig.~\ref{fig:avalanche_size}) and site occupations (Fig.~\ref{fig:height_dist_rand} and Fig.~\ref{fig:height_dist_3}). Yet, considering the transient phase, the two different drives of the systems show very different behavior.

\begin{figure}
\centering\includegraphics[width=0.5\textwidth]{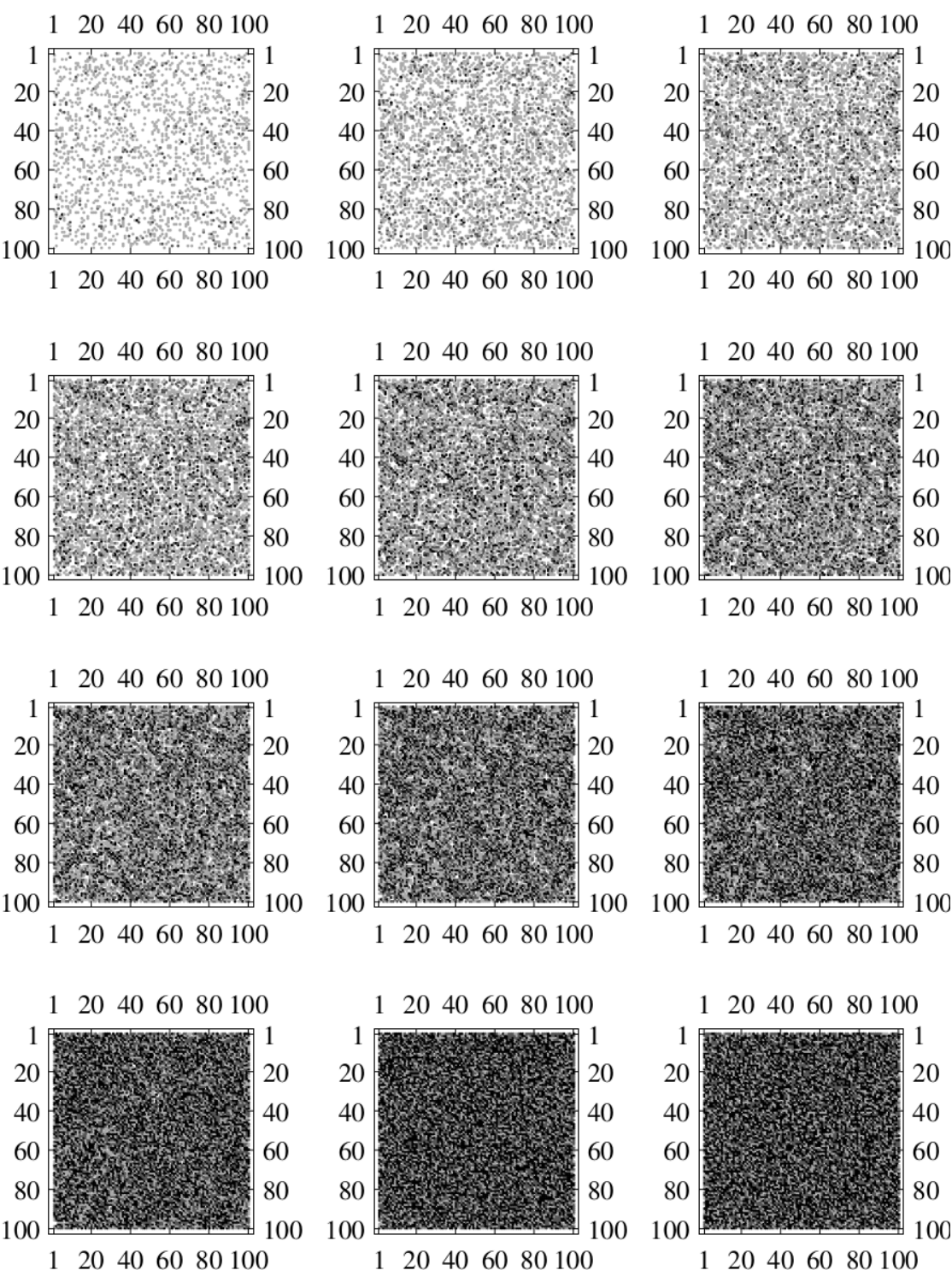}
\caption{Time evolution of the Sandpile model with random deposition. Every pixel describes the occupation of a single site: empty fields (white), fields with one (light gray), two (gray), and three (black) grains. From upper left to lower right snapshots of the system are shown from $t=2'000$ to $t=24'000$ ($dt=2'000$).}
\label{fig:matrixPlot_rnd}
\end{figure}

\begin{figure}
\centering\includegraphics[width=0.5\textwidth]{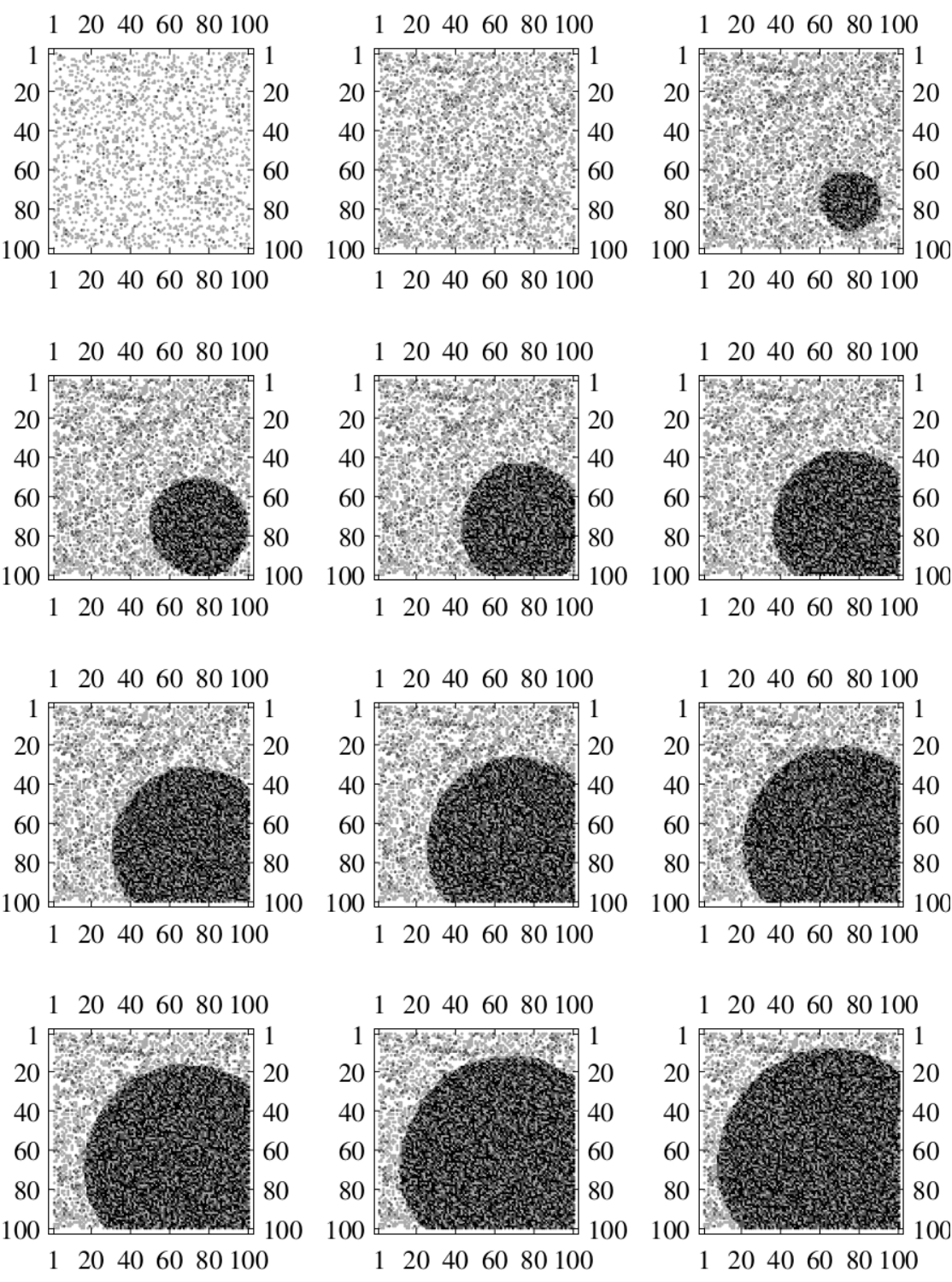}
\caption{Time evolution of the favor-3 model. Every pixel describes the occupation of a single site: empty fields (white), fields with one (light gray), two (gray), and three (black) grains. From upper left to lower right snapshots of the system are shown from $t=2'000$ to $t=24'000$ ($dt=2'000$).}
\label{fig:matrixPlot_favor3}
\end{figure}

Fig.~\ref{fig:matrixPlot_rnd} shows snapshots of the well-studied model with random deposition starting from an empty system of size $100 \times 100$. Every pixel describes the occupation of a single site. During the transient phase, the density increases homogenously over the system. The number of empty sites decreases monotonically, initially leading to a monotonic increase of fields with height one, height two and height three (see Fig.~\ref{fig:height_dist_rand}).

Fig.~\ref{fig:matrixPlot_favor3} illustrates the transient phase for the favor-3 drive. Like for the case with random deposition, initially the density increases homogenously: empty sites are transformed into sites with height one and the latter eventually become occupied by two grains. Yet, as soon as the first field with three grains appears this leads immediately to the launch of the first avalanche, whereas in the random case this happens only with probability $\frac{1}{\text{system size}}$. Therefore, the generation of every site of height three yields a small density-increase in its neighborhood. However, with this higher density, also the probability of generating the next site of height three, and thus the next avalanche, increases as compared to the rest of the system. Hence, the first few avalanches serve as condensation nuclei. Contrary to the ordinary Sandpile model, we do not obtain a homogeneous growth of density, but a cohesive, circular, high-density area which successively spreads over 
system 
(see Fig.~\ref{fig:matrixPlot_favor3}). Once the majority of all fields is covered, the average occupation saturates to similar values as for the random case (see Eq.~\eqref{eq:siteOccRnd} and Eq.~\eqref{eq:siteOccf3}).

\begin{figure}
\centering\includegraphics[width=0.5\textwidth]{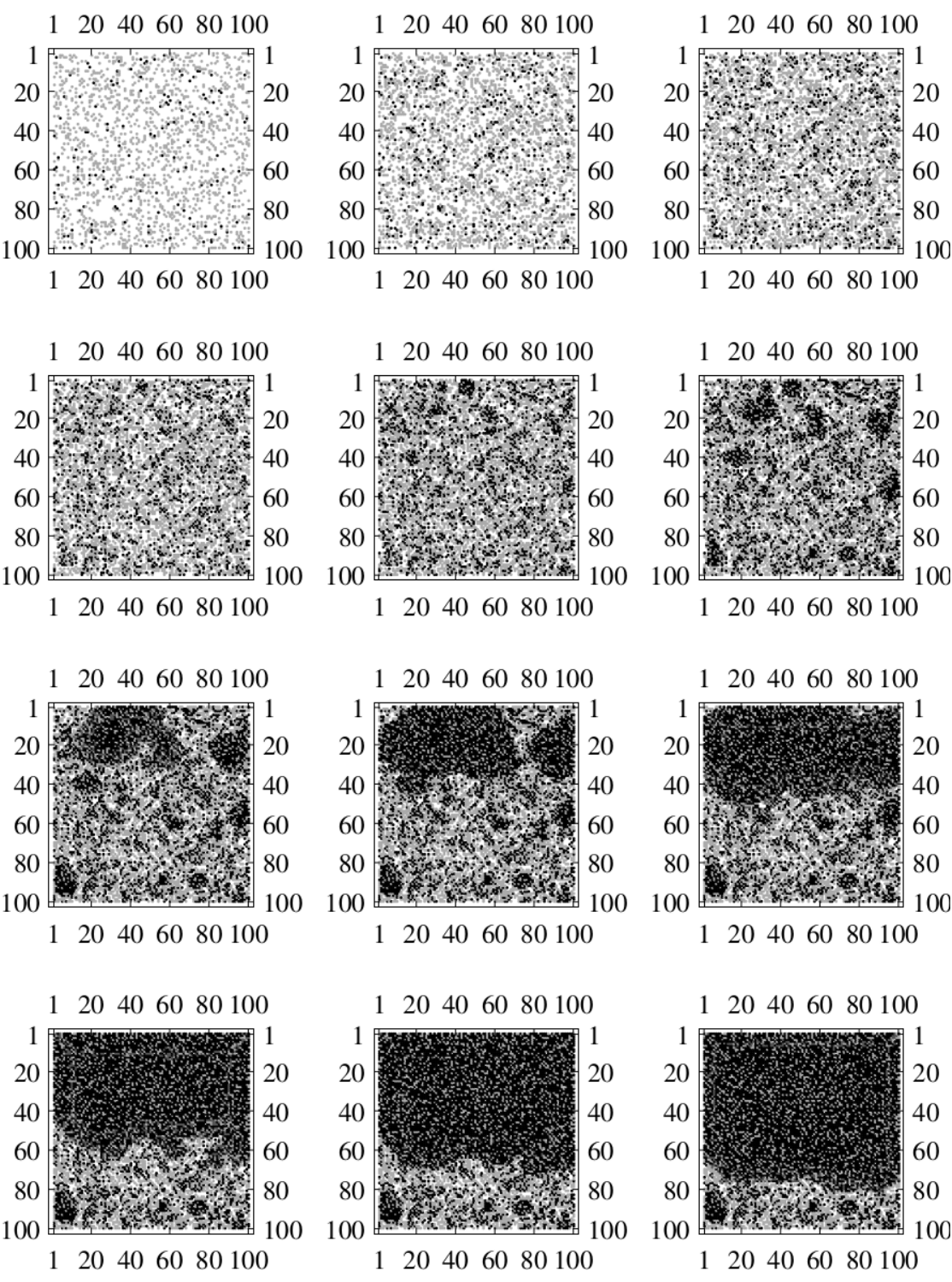}
\caption{Time evolution of the favor-2 model. Every pixel describes the occupation of a single site: empty fields (white), fields with one (light gray), two (gray), and three (black) grains. From upper left to lower right snapshots of the system are shown from $t=2'000$ to $t=24'000$ ($dt=2'000$).}
\label{fig:matrixPlot_favor2}
\end{figure}

The transient phase of the favor-2-driven model is displayed in Fig.~\ref{fig:matrixPlot_favor2}. As for the random- and the favor-3 case, initially the number of empty sites decreases as fast as the number of fields with height one increases, leading to a homogenous growth of the density in the system. However, as soon as the first sites of height two appear, due to the drive, they are immediately grown to height three, leading to a density-increase. Yet, a site with an occupation of three is still stable. Hence, the chance of the emergence of multiple clusters with a higher density is more likely than in the favor-3 case. Moreover, these clusters do not have a circular shape but grow more irregularly. During the transient period, the clusters grow. In the high-density clusters, the chance of generating large avalanches is again higher than in the rest of the system and therefore also the generation of additional sites with height two. The latter then trigger again an additional increase of density. 
Eventually the clusters merge to larger ones until the giant component covers the whole system.

The favor-1 and favor-0 cases evolve strongly distinct from the two models with other drives already from the beginning. Compared to them the density successively increases more than less homogenously over the system. The favoring process affects the system in such a way that immediately all fields which become occupied get increased to a height of two.

In the favor-0 model, due to the preferential selection, during the first $10^4$ time steps, all empty sites become occupied by a single grain. Afterwards, further depositions are processed randomly except for the fact that empty sites which occur after an avalanche are filled up right with the very next grain. At first glance, this may seem like the random model where sites topple at a height of three rather than four. It shall be emphasized that this is not the case, since in such a model the rotation symmetry of the square lattice would be broken.

In summary, during the transition process from the empty lattice to the steady state with - on average - constant density, in the originial Sandpile model the density grows homogeneously over the lattice. This also holds for the favor-0 and favor-1-driven models. Although the favor-3 model has a very similar avalanche-size distribution as the original Sandpile model and also the site-occupation distributions strongly resemble each other, its transition process is very different. Starting from a condensation necleus, the density grows cyclic around it. Similarly, for the favor-2 model, the density increase occurs heterogenously in space via the aggregation of intermediate clusters to a giant component which, eventually, spreads over the entire lattice.

\section{Summary}

Over the last decades SOC has been studied intensively. Numerous variants of the original model of Bak, Tang and Wiesenfeld have been investigated in order to clarify the universality of SOC. However, the influence of the drive has widely been neglected so far.

In this paper we have examined the impact of a state-dependent drive in the Abelian Sandpile model of Dhar. As in the original model, we still chose sites randomly, but now we only increased their local height if it had some specific value.

For a preference of sites with height three (favor-3) the system remains in a critical state. We found the same critical exponent for both the regular Sandpile model with random deposition and the favor-3 model: $\alpha_\text{random} \simeq \alpha_\text{favor3} \simeq 1.021 \pm 0.033$. In contrast, the other drives seem to destroy criticality. The favor-2 model does not show any distribution of small avalanches: all avalanches cover basically the entire system. By preparing systems with a fraction of sites with height three we could relate this observation to a percolation-like phenomenon.

Furthermore, we investigated the transient processes of the different models. Despite the similarity of the distribution of avalanches in the stationary state between random and favor-3 deposition, the transient approach to criticality differs strongly. In the random model both the density of grains and the appearance of avalanches is homogenous over the system. In contrast, in the favor-3 case early random density fluctuations lead to the emergence of a condensation nucleus with high-density which eventually spreads over the system.

\begin{acknowledgments}
The authors thank Barbara Drossel and Andreas Engel for helpful discussions, as well as Haye Hinrichsen for scientific advice. 
\end{acknowledgments}

M.W. and J.F. contributed equally to this work.

\appendix

\section{Transient of site occupations for the random-deposition model} \label{sec:siteOccTransientRnd}

The shape of the curves in Fig.~\ref{fig:height_dist_rand} during the transient can be estimated by considering the population dynamics of the occupation numbers in a mean field-like approach:
We first neglect any interaction between sites, i.e. we neglect Eq.~\eqref{eq:avalanches} and only consider the deposition of particles without toppling. We denote the number of sites with height $h$ by $N_h$ and the total number of sites by $N$. After starting the deposition of grains $N_0$ decreases, since particles are deposited. The probability to hit a site of height~$0$ and grow it to height~$1$ is $p_{0\rightarrow 1} = \frac{N_0}{N} $. This means the evolution of $N_0$ is:
Solving this equation with the initial value of $N_0 = N$ gives:
\begin{align}
 N_0(t) = N e^{-\frac{t}{N}} 
\end{align}
Analogously, we define $p_{1\rightarrow2} = \frac{N_1}{N}$ and $p_{2\rightarrow3} = \frac{N_2}{N}$. It is straightforward that the corresponding equations for the other heights then read:
\begin{align}
 \frac{\mathrm{d}N_1(t)}{\mathrm{d}t} &= p_{0\to1} -p_{1\rightarrow 2} = e^{-\frac{t}{N}} -\frac{N_1(t)}{N} \\
 \frac{\mathrm{d}N_2(t)}{\mathrm{d}t} &= p_{1\to2} -p_{2\rightarrow 3} \\
 \frac{\mathrm{d}N_3(t)}{\mathrm{d}t} &= p_{2\to3}
\end{align}
Solving this set of equations with the initial values $N_1\left(0\right) = N_2\left(0\right) = N_3\left(0\right) = 0$ yields:
\begin{equation}
 \begin{split}
 N_1(t) &= e^{-\frac{t}{N}}t \\
 N_2(t) &= \frac{e^{-\frac{t}{N}}t^2}{2N} \\
 N_3(t) &= \frac{e^{-\frac{t}{N}} \left(2 N^2  e^{\frac{t}{N}}-2 N^2-2 N t-t^2\right)}{2 N}
 \end{split}
 \label{eq:n123simple}
\end{equation}
\begin{figure}
\centering\includegraphics[width=0.5\textwidth]{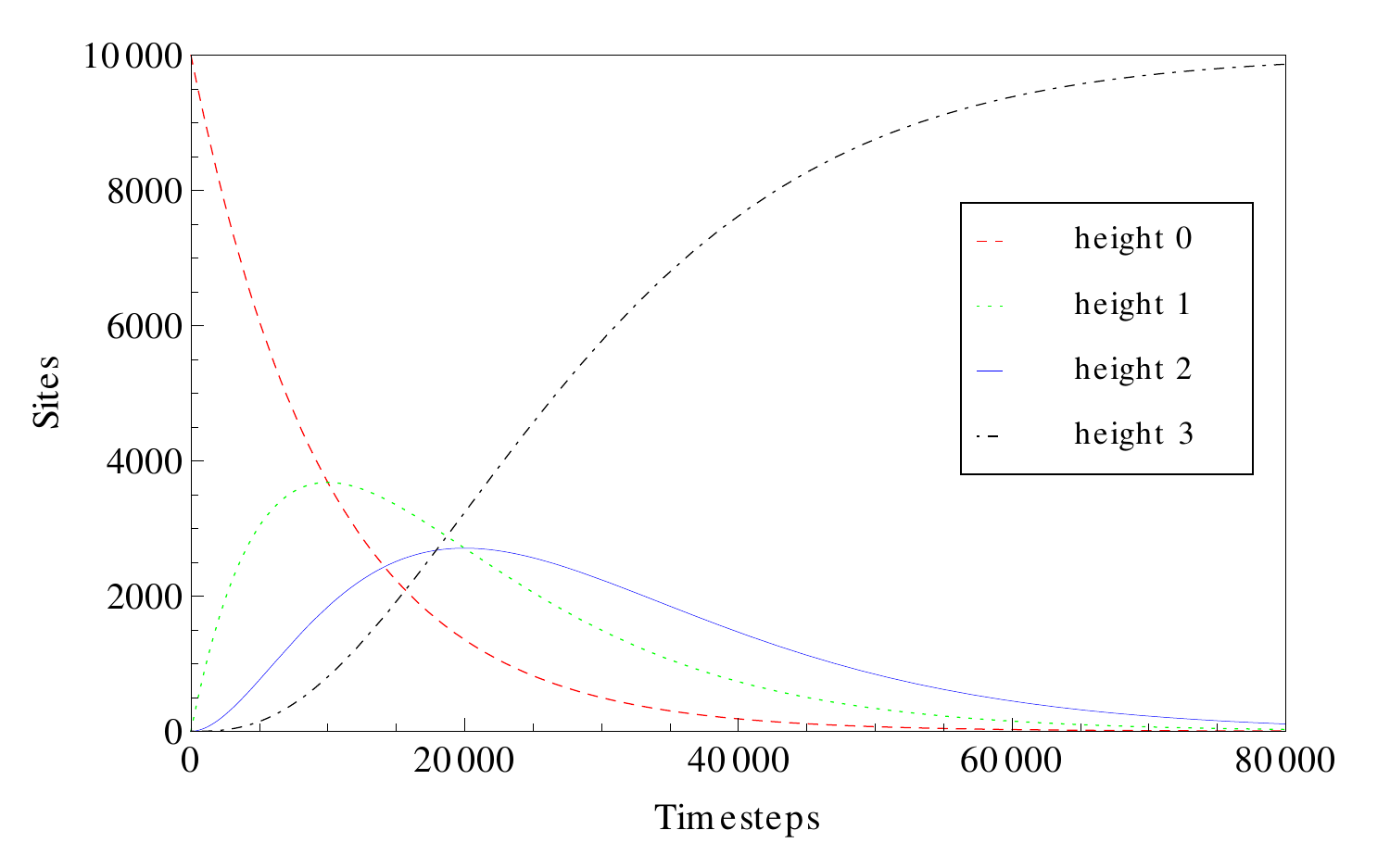}
\caption{(Color online) Plot of the analytic solution of Eqs.~\eqref{eq:n123simple}.}
\label{fig:transient_simple_dgl}
\end{figure}
\begin{figure}
\centering\includegraphics[width=0.5\textwidth]{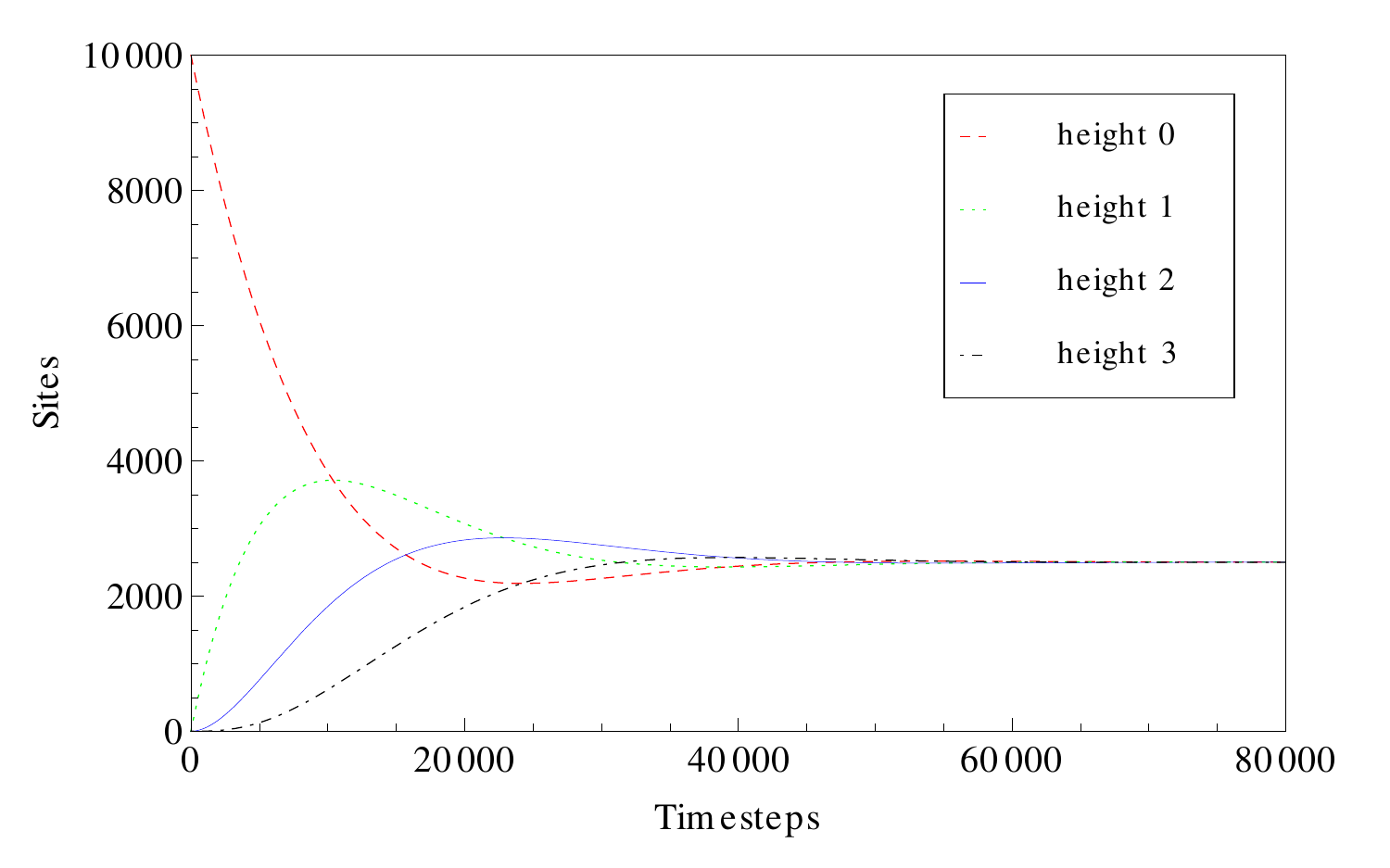}
\caption{(Color online) Plot of the analytic solution of Eqs.~\eqref{eq:mf2}.}
\label{fig:transient_better_dgl}
\end{figure}
Plotting the number of sites versus time for a system of size $N = 100 \times 100$ already resembles the transient of the random-deposition model up to a time of about $15'000$ (see Fig.~\ref{fig:transient_simple_dgl}). Since we neglect any toppling, at infinite time all sites have height three.
Thus, we introduce the equations that consider the toppling of sites with an occupation higher than three. We add an additional term to Equations~\eqref{eq:n123simple} for sites with height three and zero. Doing so, we get a system of four coupled differential equations. Since the number of particles in our system without dissipation is conserved, it is
\begin{equation}
N_1 + N_2 + N_3 + N_0 = N
\end{equation}
and hence
\begin{equation}
\frac{\mathrm{d}N_1(t)}{\mathrm{d}t} + \frac{\mathrm{d}N_2(t)}{\mathrm{d}t} + \frac{\mathrm{d}N_3(t)}{\mathrm{d}t} + \frac{\mathrm{d}N_0(t)}{\mathrm{d}t} = 0.
\end{equation}
This means we can define the last equation for $N_3$ as:
\begin{equation}
\frac{\mathrm{d}N_3(t)}{\mathrm{d}t} = - \frac{\mathrm{d}N_2(t)}{\mathrm{d}t} - \frac{\mathrm{d}N_1(t)}{\mathrm{d}t} - \frac{\mathrm{d}N_0(t)}{\mathrm{d}t}
\end{equation}
Now our four equations with added toppling and $p_{3\rightarrow0} = \frac{N_3}{N}$ read:
\begin{equation}
 \begin{split}
 \frac{\mathrm{d}N_0(t)}{\mathrm{d}t} &= -p_{0\to 1} + p_{3\to0} \\
 \frac{\mathrm{d}N_1(t)}{\mathrm{d}t} &= p_{0\to1} -p_{1\rightarrow 2} \\
 \frac{\mathrm{d}N_2(t)}{\mathrm{d}t} &= p_{1\to2} -p_{2\rightarrow 3} \\
 \frac{\mathrm{d}N_3(t)}{\mathrm{d}t} &= - \frac{\mathrm{d}N_2(t)}{\mathrm{d}t} - \frac{\mathrm{d}N_1(t)}{\mathrm{d}t} - \frac{\mathrm{d}N_0(t)}{\mathrm{d}t}
\end{split}
\label{eq:mf2}
\end{equation}
Since these equations do not consider any correlation between the sites, the evaluation leads to a mean-field result where, at infinite time, all heights occur with same probability (Fig.~\ref{fig:transient_better_dgl}):
\begin{equation}
 \begin{split}
N_0(t) &= \frac{1}{4} N   e^{-\frac{2 t}{N}} \left(e^{\frac{2 t}{N}}+2 e^{\frac{t}{N}} \cos   \left(\frac{t}{N}\right)+1\right)\\
 N_1(t) &=  \frac{1}{4} N e^{-\frac{2 t}{N}}   \left(e^{\frac{2 t}{N}}+2 e^{\frac{t}{N}} \sin   \left(\frac{t}{N}\right)-1\right)\\
   N_2(t) &= \frac{1}{4} N   e^{-\frac{2 t}{N}} \left(e^{\frac{2 t}{N}}-2 e^{\frac{t}{N}} \cos   \left(\frac{t}{N}\right)+1\right)\\
   N_3(t) &= \frac{1}{4} N   e^{-\frac{2 t}{N}} \left(e^{\frac{2 t}{N}}-2 e^{\frac{t}{N}} \sin   \left(\frac{t}{N}\right)-1\right)
\end{split}
\end{equation}
\begin{figure}
\centering\includegraphics[width=0.5\textwidth]{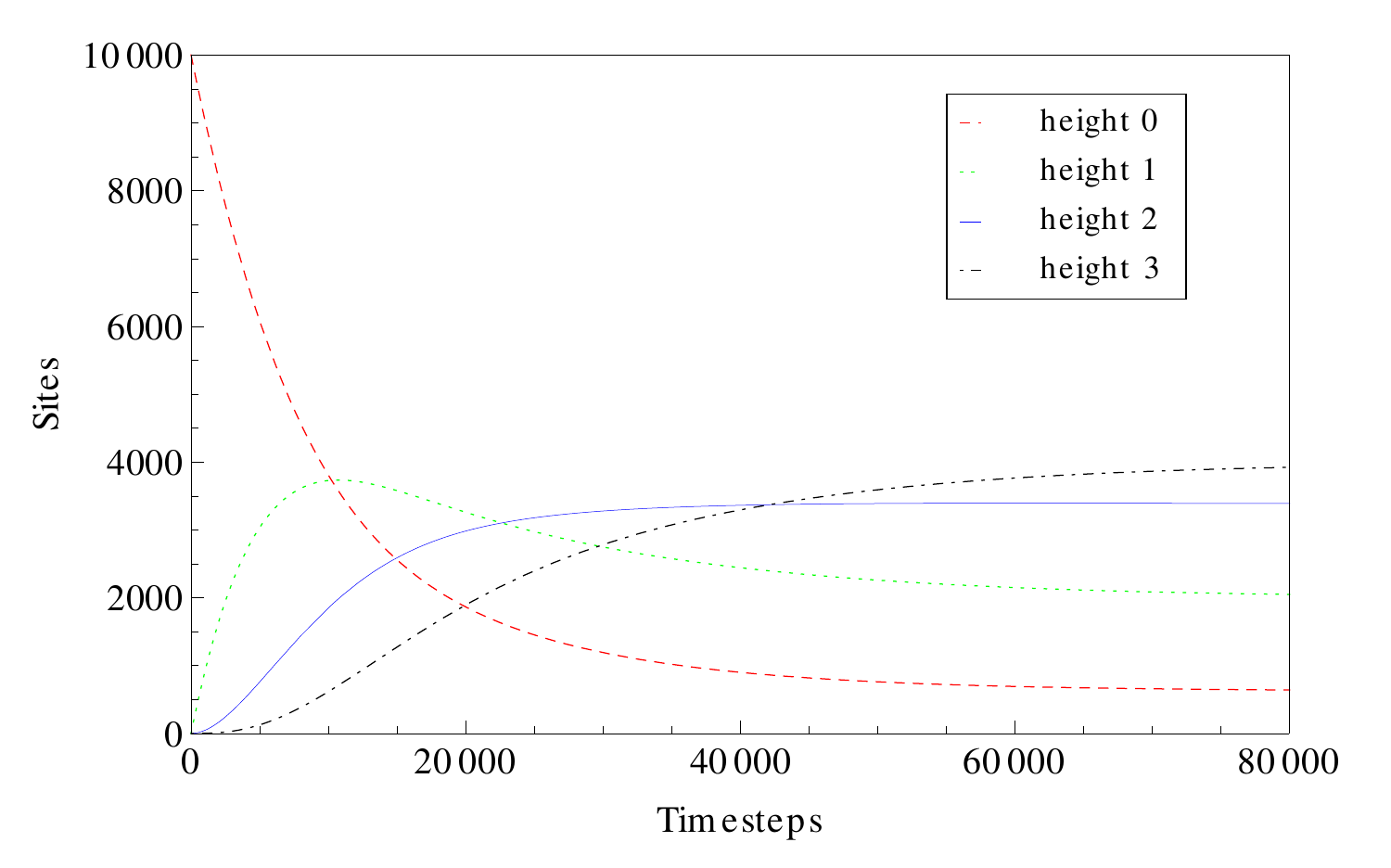}
\caption{(Color online) Plot of the analytic solution of Eqs.~\eqref{eq:mf3_1} and~\eqref{eq:mf3_2}.}
\label{fig:transient_best_dgl}
\end{figure}
To account for correlations between the fields, we now add the first order of avalanches. This means, a deposition on a site of height three now only increases the number of sites with height zero if none of the the field's nearest neighbors has height three, or all of them have height three:
\begin{equation}
 \frac{dN_0(t)}{dt} = -p_{0\to 1} + p_{3\to0} \left(\left( 1- p_{3\to0} \right)^4 + p_{3\to0}^4 \right)
 \label{eq:mf3_1}
\end{equation}

Analogously, sites of height one can be created, if a site of height three topples and there is exactly one nearest neighbor of height three that topples in the following avalanche. Likewise, sites of height two can be created, if a site of height three topples and there are exactly two next neighbors of height three that topple subsequently:
\begin{equation}
 \begin{split}
 \frac{\mathrm{d}N_1(t)}{\mathrm{d}t} = 
  -p_{1\to 2} &+ p_{0\to1}\\ &+ 4 \cdot p_{3\to0} \left(\left( 1- p_{3\to0} \right)^3 p_{3\to0} \right) \\
 \frac{\mathrm{d}N_2(t)}{\mathrm{d}t} = 
  -p_{2\to 3} &+ p_{1\to2} \\&+ 6 \cdot p_{3\to0} \left(\left( 1- p_{3\to0} \right)^2 p_{3\to0}^2 \right) 
\end{split}
\label{eq:mf3_2}
\end{equation}
The factors $4$ and $6$ follow from the corresponding binomial coefficients. To our knowledge there is no analytic solution for these modified equations. The numerical solution in Fig.~\ref{fig:transient_best_dgl} shows that this first order approximation already describes the actual time evolution (Fig.~\ref{fig:height_dist_rand}) very well.

Since these approximations only consider the implications of avalanches triggered at the four nearest neighbors, it takes longer for the system to reach the steady state. While in the simulations, after approximately $2.2 \times 10^4$ time steps, the system reaches its steady state with saturating density, in the first-order approximation, it takes almost $8 \times 10^4$ time steps until the density saturates at a mean height of $2$.

\bibliographystyle{apsrev4-1}

%


\end{document}